\documentstyle[12pt,epsfig]{article}        % Simple LaTeX, use with
\newcommand{\gsim}{\raisebox{-0.07cm}{$\,\stackrel{>}{{\scriptstyle
 \sim}}\, $} }
\newcommand{\lsim}{\raisebox{-0.07cm}{$\,\stackrel{<}{{\scriptstyle
 \sim}}\, $} }

\newcommand\beq{\begin{equation}}
\newcommand\eeq{\end{equation}}
\newcommand\bea{\begin{eqnarray}}
\newcommand\eea{\end{eqnarray}}

\newlength{\dinwidth}                       
\newlength{\dinmargin}                      
\def\gev{\rm GeV}
\def\lsim{\mathrel{\rlap{\lower4pt\hbox{\hskip1pt$\sim$}}
    \raise1pt\hbox{$<$}}}                % less than or approx. symbol
\def\gsim{\mathrel{\rlap{\lower4pt\hbox{\hskip1pt$\sim$}}
    \raise1pt\hbox{$>$}}}                % greater than or approx. symbol
\title{$\!\! The Heavy-Flavour Contribution to Proton Structure\!\!\!\! $}
\author
{K. Daum, J. Bl\"umlein$^{\,\rm a}$, S. Riemersma
 \address{DESY--Zeuthen, Platanenallee 6, D--15735 Zeuthen, Germany},
and A. Vogt
 \address{Institut f\"ur Theoretische Physik, Universit\"at W\"urzburg,
          Am Hubland, D--97074 W\"urzburg, Germany}
}

\begin{document}
% Begin of extra titlepage for the preprint version
\begin{titlepage}

\begin{flushleft}
DESY 96--205 \\[0.1cm] FSU-HEP-960917 \\[0.1cm] CERN-TH/96-262    \\[0.1cm]
September 1996 
\end{flushleft}
\vspace{0.4cm}
\begin{center}
\large
{\bf The Heavy-Flavour Contribution to Proton Structure} \\
\vspace{0.7cm}
%\Large
K. Daum\\
\vspace{0.2cm}
%\large 
{\it
Rechenzentrum, Universit\"at Wuppertal\\
\vspace{0.1cm}
Gau\ss{}stra\ss{}e 20, D-42097 Wuppertal, Germany}\\
\vspace{0.3cm}
%\Large
S. Riemersma\\
\vspace{0.2cm}
%\large 
{\it
DESY--Zeuthen \\
\vspace{0.1cm}
Platanenallee 6, D--15735 Zeuthen, Germany }\\
\vspace{0.3cm}
%\Large
B.W. Harris\\
\vspace{0.2cm}
%\large 
{\it
Department of Physics, Florida State University\\
\vspace{0.1cm}
Tallahassee, Florida, 32306-3016, USA}\\
\vspace{0.3cm}
%\Large
E. Laenen\\
\vspace{0.2cm}
%\large 
{\it
CERN Theory Division\\
\vspace{0.1cm}
CH-1211 Geneva 23, Switzerland}\\
\vspace{0.3cm}
%\Large
J. Smith\\
\vspace{0.2cm}
%\large 
{\it
Institute for Theoretical Physics, SUNY Stony Brook\\
\vspace{0.1cm}
Stony Brook NY 11794-3840, USA}\\
\vspace{0.5cm}
{\bf Abstract}
\end{center}
\vspace{-0.1cm}
We present theoretical and experimental considerations pertaining to
deeply inelastic heavy-flavour production at HERA.  The various theoretical
uncertainties in the cross section 
calculation are discussed.  Cuts are imposed to determine the fraction
of charm production accessible to the detectors.  The production of
charm at asymptotic $Q^2$ and bottom production are also
covered. Experimental aspects 
include current charm production data analysis and prospects for
future analyses including anticipated high precision and
distinguishing photon-gluon fusion charm events from excitation from
the charm parton density. The feasibility of measuring 
$F_2^{b\overline b}(x,Q^2)$ is investigated.
\vfill 
\noindent
\normalsize
$^{\ast} ${\it  To appear in abbreviated form in the proceedings of the
workshop ``Future Physics at HERA'', DESY, Hamburg, 1996.}
\end{titlepage}
%\begin{document}
\vspace*{1cm}
\begin{center}  \begin{Large} \begin{bf}
The Heavy-Flavour Contribution to Proton Structure\\
  \end{bf}  \end{Large}
  \vspace*{5mm}
  \begin{large}
K. Daum$^{a,b}$, S. Riemersma$^c$, B.W. Harris$^d$, E. Laenen$^e$, 
J. Smith$^{b,f}$ 
  \end{large}
\end{center}
$^a$ Rechenzentrum, Universit\"at Wuppertal, Gau\ss{}stra\ss{}e 20,
D-42097 Wuppertal, Germany\\
$^b$ DESY, Notkestrasse~85,~D-22603~Hamburg,~Germany\\
$^c$ DESY-Zeuthen, Platanenallee 6 D-15738 Zeuthen, Germany\\
$^d$ Department of Physics, Florida State University,
     Tallahassee, Florida, 32306-3016, USA\\
$^e$ CERN Theory Division, CH-1211 Geneva 23, Switzerland\\
$^f$ Institute for Theoretical Physics, SUNY Stony Brook, NY 11794-3840, USA\\
\begin{quotation}
\noindent
{\bf Abstract:}
We present theoretical and experimental considerations pertaining to
deeply inelastic heavy-flavour production.  The various theoretical
uncertainties in the cross section 
calculation are discussed.  Cuts are imposed to determine the fraction
of charm production accessible to the detectors.  The production of
charm at asymptotic $Q^2$ and bottom production are also
covered. Experimental aspects 
include current charm production data analysis and prospects for
future analyses including anticipated high precision and
distinguishing photon-gluon fusion charm events from excitation from
the charm parton density. The feasibility of measuring 
$F_2^{b\overline b}(x,Q^2)$ is investigated.
\end{quotation}
\section{Introduction}

Heavy-flavour production in deeply inelastic scattering (DIS) at HERA
is now emerging as a very important means of studying proton
structure.  The ink is still drying on the first experimental reports
of charm production from photon-mediated DIS at HERA \cite{H1,ZEUS}.
The next-to-leading order (NLO) calculations have also
been published within the last four years. The inclusive calculation
of the photon-mediated heavy-flavour structure functions 
$F_{2,L}^{{\rm hq}}(x,Q^2,m^2)$ \cite{lrsn1}, the inclusive single
differential distribution $dF_{2,L}^{{\rm hq}}/dO $ \cite{lrsn2} ($O$ being the
transverse momentum $p_t$ of the heavy quark and the rapidity $y$),
and the fully differential calculation \cite{hs} are 
now available for a complete NLO analysis of the photon-mediated
heavy-flavour structure function.  

In section \ref{sec:tha}, following a short review of the necessary
formulae, we investigate various theoretical issues surrounding DIS
heavy-flavour 
production.  The primary sources of theoretical
uncertainty include the imprecisely determined charm quark mass and
renormalization and factorization scale dependences.  Additional
impediments to a clean extraction of the gluon density from
heavy-flavour production include the effects of light-quark ($u,d,s$)
initiated heavy-flavour production and the influence of the
longitudinal heavy-flavour structure function $F_L^{{\rm
  hq}}(x,Q^2,m^2)$ upon the cross section results.  

We investigate the effect of realistic cuts in $p_t$ and
pseudorapidity $\eta$ on the cross section and determine acceptance
probabilities as a function of $x$ and $Q^2$.  Charm production in the
limit $Q^2 \gg m_c^2 $ \cite{bmmsv} and the transition of charm
production from boson-gluon 
fusion at low $Q^2$ to excitation from the charm density as $Q^2$
becomes much larger than $m_c^2$ is then discussed, followed by a
cursory view of bottom production.

In section
\ref{sec:ea},  we discuss the analysis of the 1994 HERA data.
Identification methods of the produced $D^0,D^{*\pm}$ are outlined.
The measured cross section \cite{H1} is compared with various
theoretical predictions.    
A determination of the source of the charm production is performed,
revealing the primary production mechanism at presently measured $Q^2$
values is photon-gluon fusion rather than stemming from the charm
parton density.  The charm structure function $F_2^{c \overline c}$ is
extracted and the ratio $F_2^{c \overline c}/F_2$ is determined.

Future experimental prospects include the installation of silicon
vertex detectors, enabling greater charm and bottom hadron detection
efficiency.  The anticipated luminosity of $500\, pb^{-1}$ will allow
detailed studies of charm production dynamics.  The transition from
boson-gluon fusion of charm to excitation from the charm quark sea
should become apparent as the accessible $Q^2$ grows.
The predicted bottom quark production cross section will enable
studies of
$F_2^{b \overline b}/F_2^{c \overline c}$ as a function of $x$ and $Q^2$ with
reasonable precision.

\section{Theoretical Aspects}
\label{sec:tha}

\subsection{Background}
\label{ssec:bg}

The reaction under study is
\begin{equation}
e^-(l) + P(p) \rightarrow e^-(l') + 
Q(p_1)(\bar{Q}(p_1))+X\,,
\label{eq:prod}
\end{equation}
where $P(p)$ is a proton with momentum $p$, $Q(p_1)(\bar{Q}(p_1))$ 
is a heavy (anti)-quark with momentum $p_1$ ($p_1^2 = m^2$) and
$X$ is any hadronic state allowed by quantum number conservation. The
cross section may be expressed as
\begin{equation}
\label{eq:sigma}
\frac{d^2\sigma}{dx \, dQ^2} = 
\frac{2 \pi \alpha^2}{x\,Q^4} \left[ ( 1 + (1-y)^2 )
F_2^{{\rm hq}}(x,Q^2,m^2) -y^2
F_L^{{\rm hq}}(x,Q^2,m^2) \right]\,,
\end{equation}
where 
\begin{equation}
q = l - l'\,, \qquad Q^2 = -q^2\,,\qquad x = \frac{Q^2}{2p\cdot q} 
\,, \qquad y = \frac{p\cdot q}{p \cdot l}\,. \label{four}
\end{equation}
The inclusive structure functions $F_{2,L}^{{\rm hq}}$ were calculated 
to next-to-leading order (NLO)
in Ref.~\cite{lrsn1}. The results can be written as
\begin{eqnarray}
F_{k}(x,Q^2,m^2) &=&
\frac{Q^2 \alpha_s(\mu^2)}{4\pi^2 m^2}
\int_x^{z_{\rm max}} \frac{dz}{z}  \Big[ \,e_H^2 f_g(\frac{x}{z},\mu^2)
 c^{(0)}_{k,g} \,\Big] \nonumber \\
&+&\frac{Q^2 \alpha_s^2(\mu^2)}{\pi m^2}
\int_x^{z_{\rm max}} \frac{dz}{z}  \bigg \{ \,e_H^2 f_g(\frac{x}{z},\mu^2)
 (c^{(1)}_{k,g} + \bar c^{(1)}_{k,g} \ln \frac{\mu^2}{m^2})
 \nonumber \\
&+&\sum_{i=q,\bar q} \Big[ e_H^2\,f_i(\frac{x}{z},\mu^2)
 (c^{(1)}_{k,i} + \bar c^{(1)}_{k,i} \ln \frac{\mu^2}{m^2})  
+ e^2_{L,i}\, f_i(\frac{x}{z},\mu^2) d^{(1)}_{k,i}  \, \Big]  \,\bigg \} \,,
\label{eq:strfns}
\end{eqnarray}
where $k = 2,L$ and the upper boundary on the integration is given by
$z_{\rm max} = Q^2/(Q^2+4m^2)$. The functions $f_i(x,\mu^2)\,, (i=g,q,\bar q)$
denote the parton densities in the proton and $\mu$ stands for the
mass factorization scale
which has been set equal to the renormalization scale. The 
$c^{(l)}_{k,i}(\zeta, \xi)\,(i=g\,,q\,,\bar q\,;l=0,1),\,
\bar c^{(l)}_{k,i}(\zeta, \xi) \, (i=g,q,\bar q;l=1),$ and
$d^{(l)}_{k,i}(\zeta, 
\xi) \, (i=q\,,\bar q\,;l=1)$
are coefficient functions and are represented in the
$\overline{\rm MS}$ scheme.
They depend on the scaling variables $\zeta$ and $\xi$ defined by
%----(6)
\begin{equation}
\zeta = \frac{s}{4m^2} - 1\quad  \qquad \xi = \frac{Q^2}{m^2}\,.
\end{equation}
where $s$ is the square of the c.m. energy of the
virtual photon-parton subprocess $Q^2(1-z)/z$.
In Eq.~(\ref{eq:strfns}) we 
distinguish between the coefficient functions with respect to their origin.
The coefficient functions indicated by
$c^{(l)}_{k,i}(\zeta, \xi),\bar c^{(l)}_{k,i}(\zeta, \xi)$
originate from the partonic
subprocesses where the virtual photon is coupled to the heavy quark, whereas
$d^{(l)}_{k,i}(\zeta, \xi)$
comes from the subprocess where the virtual photon interacts with the
light quark. 
The former are multiplied by the charge squared
of the heavy quark $e_H^2$, and the latter 
by the charge squared of the light quark $e_L^2$ respectively
(both in units of $e$).
Terms proportional to $e_H e_L$ integrate
to zero for the inclusive structure functions.
Furthermore we have isolated the factorization scale dependent logarithm
$\ln(\mu^2/m^2)$.

The scale dependence and the poorly known charm quark mass are the 
largest contributors to the theoretical uncertainty.  The effects of
$F^{{\rm hq}}_L$ and the light-quark initiated contributions are
discovered also to be important in the analysis.  To aid the
experimental analysis, the fully differential program \cite{hs} is
used to apply a series of cuts to determine the percentage of events the
detectors are likely to see in bins of $x$ and $Q^2$.  With the
planned inclusion of silicon vertex detectors, the ability to see
bottom events increases dramatically, motivating the presentation of
results for the cross section and $F_2^{b {\overline
    b}}(x,Q^2,m_b^2)$.  HERA is in a unique position to evaluate the
transition of charm production from 
photon-gluon fusion to excitation from the charm parton density.
For other phenomenological investigations, see \cite{lrsn3,av,lbhmmrsv}.

\subsection{Code Update}

For this study we use an updated version of the code based upon
\cite{rsn}.  The original code was based upon fitting the coefficient
functions described in eq. (\ref{eq:strfns}) using a two-dimensional tabular
array of points in $\zeta$ and $\xi$.  The coefficient functions
were generated via a linear interpolation between the calculated
points.  The linear interpolation was insufficiently accurate and
required a more sophisticated interpolation procedure.

The present interpolation procedure is based upon a Lagrange
three-point interpolation formula, see eq. 25.2.11 \cite{as}.  The
results have been thoroughly compared with the original code in
\cite{lrsn1} and \cite{hs} and excellent agreement has been 
established.\footnote{The code is available at
  http://www.ifh.de/theory/publist.html. }

\begin{figure} 
\centering
\epsfig{file=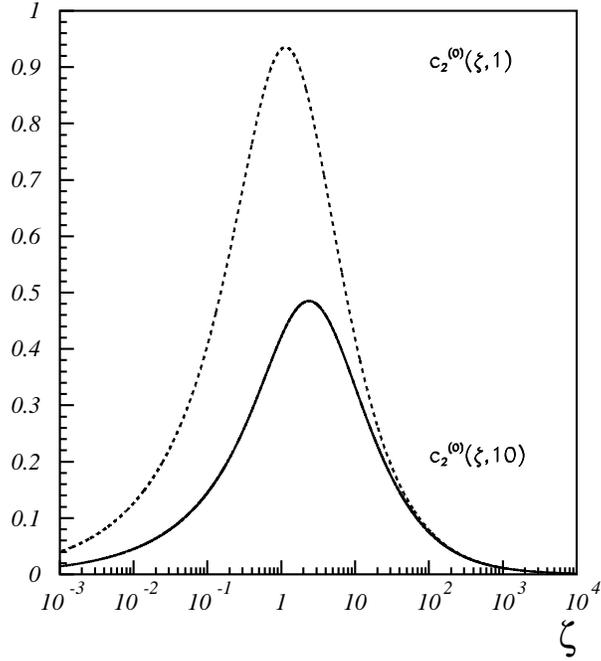,width=10cm,angle=0}
\caption{\label{fig:born}{\it
The Born coefficient function $c^{(0)}_2(\zeta, \xi)$, for $\xi =
1,10$.}}
\end{figure}

\begin{figure} 
\centering
\epsfig{file=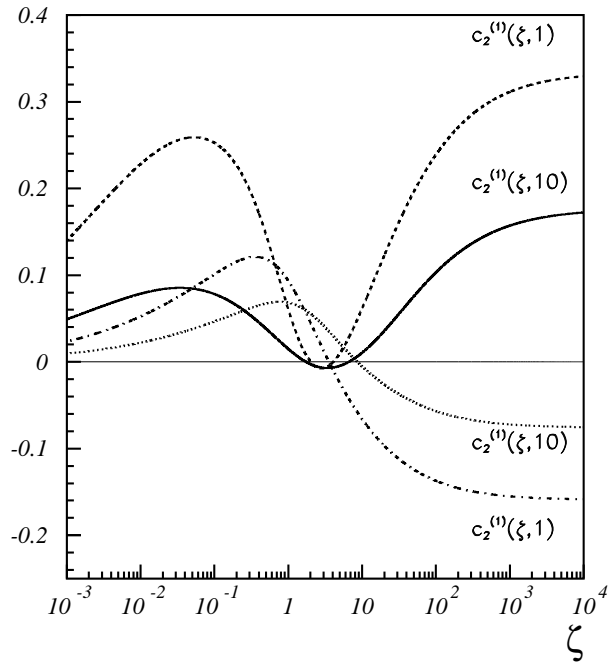,width=10cm,angle=0}
\caption{\label{fig:corr}{\it The NLO coefficient functions
    $c^{(1)}_{2,g}(\zeta, \xi)$, for $\xi = 1,10$ (upper curves at
    large $\zeta$) and ${\protect\overline c}^{(1)}_{2,g}(\zeta,\xi)$
    for $\xi = 1,10$ (lower curves at large $\zeta$).}} 
\end{figure}

As a demonstration of the code, we present results for the Born
$c^{(0)}_{2}(\zeta, \xi = 1,10)$ in Fig. \ref{fig:born} and
$c^{(1)}_{2,g}(\zeta, \xi = 1,10)$ and ${\overline
  c}^{(1)}_{2,g}(\zeta, \xi = 1,10)$ in Fig. \ref{fig:corr}.


\subsection{Scale and Parton Density Related Issues}

\begin{figure}[htb] 
\centering
\epsfig{file=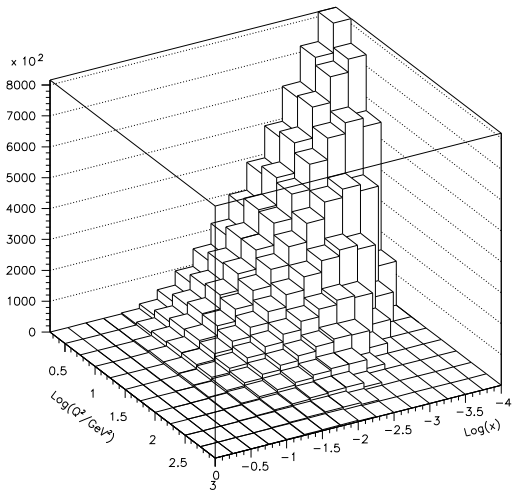,width=7.5cm,angle=0}
\caption{\label{fig:events}{\it
Projected number of DIS charm events for an integrated luminosity of
500 $pb^{-1}$ binned in $x$ and $Q^2$ with no cuts applied.
  }}
\end{figure}

The most important numerical sources of theoretical uncertainty in DIS
heavy-flavour production are the factorization/renormalization scale
dependence and the poorly known charm quark mass.  Varying $\mu$ in
Eq. (\ref{eq:strfns}) indicates the stability of the NLO result
against scale changes.  To get a concrete idea of the effect of scale
variations, we first construct a ``data'' set.  The number of DIS charm
events is calculated for an integrated luminosity of $500 \, pb^{-1}$,
using CTEQ3M \cite{cteq3} parton densities with $\Lambda_4 = 239$ MeV,
$m_c = 1.5$ GeV, and $\mu^2 = Q^2 + 4\, m_c^2$.  Unless otherwise
mentioned, CTEQ3M is used for all results.  The results in
Fig. \ref{fig:events}. are
produced for $1.8 < Q^2 < 1000$ GeV$^2$ and $10^{-4} < x < 1$ using
four bins per decade for both $Q^2$ and $x$.
%\begin{figure}[htb] 
The ``data'' are peaked strongly at small $x$ and $Q^2$.  As $Q^2$
grows, the events become more evenly distributed in $x$.  For small
$Q^2$, the number of events $N$ falls off as $x \rightarrow 1$.  At larger
$Q^2$, $N$ rises, peaks at intermediate $x$, then drops.  

\begin{figure}[b] 
\centering
\epsfig{file=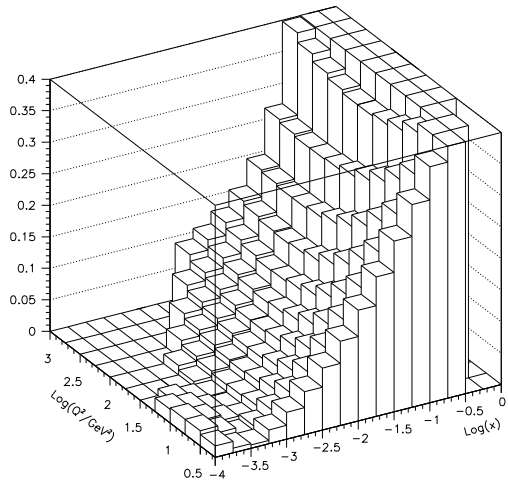,width=7.5cm,angle=0}
\caption{\label{fig:scales}
{\it
$\Delta_{\mu}(x,Q^2)$ (see Eq. (\ref{eq:delNofmu})) binned in $x$ and $Q^2$.
  }}
\end{figure}

Using the ``data'' set as a point of reference, we can investigate where in the
kinematic region the effects of scale variation are most strongly
felt.  Keeping every other parameter fixed, we determine the number of
events for $\mu^2 = 4 m_c^2$ and $\mu^2 = 4(Q^2 + 4 m_c^2)$.  We
investigate the quantity
\begin{equation}
\label{eq:delNofmu}
\Delta_{\mu}(x,Q^2) \equiv \bigg | \frac{N(\mu^2 = 4 m_c^2) - N(\mu^2
  = 4(Q^2 + 4   m_c^2))}
{2 N(\mu^2 = Q^2 + 4 m_c^2)}  \bigg |.
\end{equation}
The results are displayed in Fig.. \ref{fig:scales}.
%\begin{figure}[htb] 
Apart from the high-$x$ region, which contains very few charm events,
the scale dependence varies relatively little with 
$Q^2$.  As $x \rightarrow 0$, we find the scale dependence
disappearing.  This behaviour bodes very well for a low-$x$ extraction of the
gluon density from charm events.

\begin{figure}[t] 
\centering
\epsfig{file=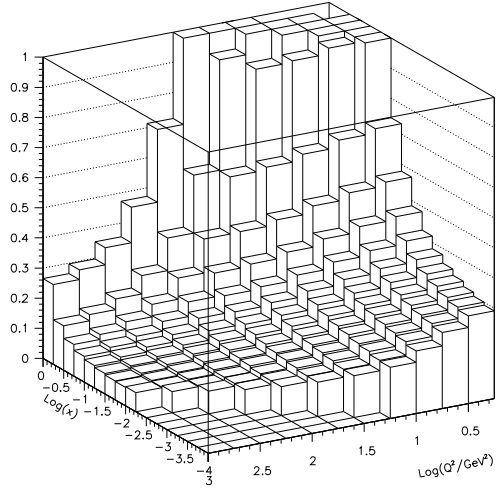,width=7.5cm,angle=0}
\caption{\label{fig:mass}
{\it
$\Delta_{m_c}(x,Q^2)$ (see Eq. (\ref{eq:delNofmc})) binned in $x$ and $Q^2$.
  }}
\end{figure}

The charm quark mass uncertainty presents a stickier problem.  A
precise measurement of the charm mass is yet to be made.  To develop
a feeling for how much of an effect the uncertainty has, we calculated
\begin{equation}
\label{eq:delNofmc}
\Delta_{m_c}(x,Q^2) \equiv \frac{N(m_c = 1.3 {\rm GeV}) - N(m_c = 1.7
  {\rm GeV})}{2 \, N(m_c = 1.5 {\rm GeV})}.
\end{equation}
Very large effects are naturally found near threshold, but as $Q^2$
increases and $x$ decreases, $\Delta_{m_c}$ approaches a value on
the order of $0.1$.  Varying the charm mass from 1.3 to 1.7
GeV is a conservative estimate;  the error induced by the uncertainty
viewed in Fig. \ref{fig:mass}. can be viewed as an upper bound.
%\begin{figure}[htb] 

\begin{figure}[tb]
\centering 
\epsfig{file=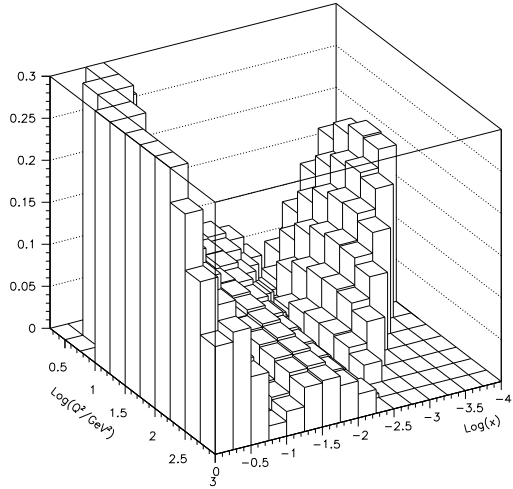,width=7.5cm,angle=0}
\caption{\label{fig:pdfs}{\it
$\Delta_{{\rm glue}}(x,Q^2)$ (see Eq. (\ref{eq:delNglue})) binned in
$x$ and $Q^2$.   }}
\end{figure}

A clear indication of the ability to extract the gluon density from
charm production is whether one can distinguish the gluon densities
from different available parton densities.  We compare the cross section
generated with CTEQ2MF\cite{cteq3} (with a flat gluon density as $x \rightarrow
0$) with GRV94HO\cite{grv94} (with a steep gluon density as $x \rightarrow 0$).
We define 
\begin{equation}
\label{eq:delNglue}
\Delta_{{\rm glue}}(x,Q^2) \equiv \bigg |\frac{N_{{\rm CTEQ2MF}} - N_{{\rm GRV94HO}}}
  {2 \, N_{{\rm CTEQ3M}}} \bigg |
\end{equation}
and show the results in Fig. \ref{fig:pdfs}.
%\begin{figure}[htb] 
Away from large $x$, $\Delta_{{\rm glue}}(x,Q^2)$ is flat as a function of $Q^2$.  In the intermediate region
in $x$, very little distinguishing power is observed.  Beginning near
$x = 10^{-3}$, a definite difference is seen.  The analysis must
seemingly extend to
$x \lsim 5 \cdot 10^{-4}$ to distinguish cleanly the gluon densities
mentioned. 

To summarize:  While the scale dependence is well under control,
the dominant source of uncertainty is clearly the charm 
quark mass, which has a strong influence on the cross section at low
$x$ and low $Q^2$.  This region is exactly the region sensitive to the
gluon density.  This strong influence of $m_c$ poses problems for a clean
extraction of the gluon density at small $x$ from inclusive measurements.

\subsection{Smaller Contributions to the Cross Section:  $F_L^{c
    {\overline c}}$ and Light-Quark Initiated Results}

\begin{figure}[tb] 
\centering
\epsfig{file=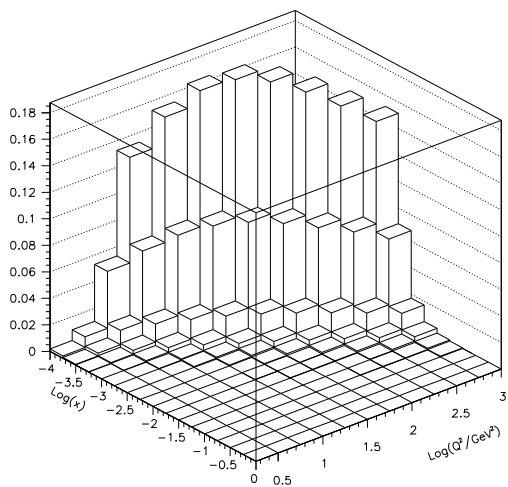,width=7.5cm,angle=0}
\caption{\label{fig:FL}
{\it
$C_{F_L}(x,Q^2)$ (see Eq. (\ref{eq:fl})) binned in $x$ and $Q^2$.
  }}
\end{figure}

A clean extraction of the gluon density may be hindered by
contributions of $F_L^{c {\overline c}}$ and light-quark initiated
contributions to the cross section.  
We investigate the fractional $F_L$ contribution to the cross section
by plotting  
\begin{equation}
\label{eq:fl}
 C_{F_L} \equiv \bigg | \frac{\sigma_{tot} - \sigma_{F_2}}{\sigma_{tot}}
 \bigg |,
\end{equation}
where $\sigma_{F_2}$ represents the contribution to the cross section
in Eq. (\ref{eq:sigma}) with $F_L$ set to zero.
The results are shown in Fig. \ref{fig:FL}.
%\begin{figure}[htb] 
We find a sizeable contribution to the cross section at high $y$.  This
overlaps with the low $Q^2$ and low $x$ region previously determined
to be deemed the most suitable for a gluon density extraction.  To do so,
however, one must take into consideration $F_L^{c {\overline c}}$.  

\begin{figure}[htb]
\centering 
\epsfig{file=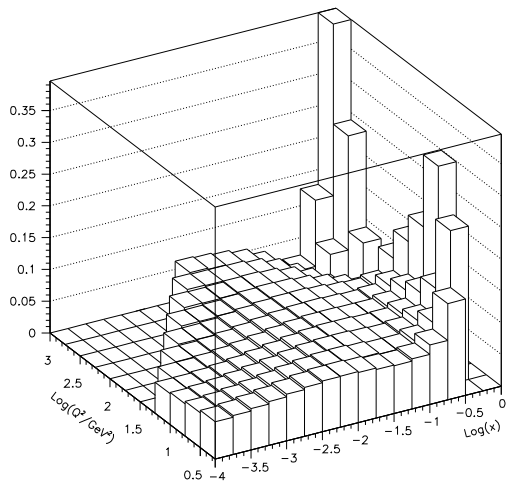,width=7.5cm,angle=0}
\caption{\label{fig:quark}
{\it
$C_{{\rm lq}}$ (see Eq. (\ref{eq:quark})) binned in $x$ and $Q^2$.
  }}
\end{figure}

We investigate the light-quark initiated contribution by determining
\begin{equation}
\label{eq:quark}
C_{{\rm lq}} =\bigg | \frac{\sigma_{{\rm lq}}}{\sigma_{tot}} \bigg |
\end{equation}
and displaying the results in Fig. \ref{fig:quark}.
The results are very nearly constant, amounting to a 5 -- 8 \%
contribution except at large $x$, where the charm contribution does
not appreciably contribute numerically.

Summarizing:  the contributions to the cross section from 
$F_L^{c {\overline c}}$ 
are noticeable in the large $y$ region, which overlaps with the small
$x$ and small $Q^2$ region, hindering the extraction of the gluon
using only $F_2^{c {\overline c}}$.  The light-quark initiated cross section
contributes on the order of 5 \% to the total cross section, and
therefore cannot be totally neglected.

\subsection{Cuts {\it vs.} No Cuts}

With the fully differential code developed in \cite{hs}, a series of
cuts can be applied determining a more realistic expectation of charm
events able to be detected.   

The cuts imposed upon the data are 
\begin{equation}
\label{eq:cuts}
|\eta_c| \leq 1.5, \,\, p_t^{c} \ge 2 \, {\rm GeV},
\end{equation}
$\eta_c$ being
the pseudorapidity of the detected charm quark.
Looking first at the low $Q^2$ range from 2 to 10 GeV$^2$, we
present the cross section with and without the aforementioned cuts in
the $x$ bins outlined in Tab.~\ref{tab:lowq2}.  These results were
generated using GRV94HO, $m_c = 1.5$ GeV, 3 light flavours, and
$\Lambda_3 = 248$ MeV.
% 2 - 10 GeV^2
\begin{table}
\begin{tabular}{||c|c||c|c|c||} \hline \hline
  $x_{min}$ & $x_{max}$ &  $\sigma$ (nb) &
  $\sigma_{cuts}$ (nb) &  Efficiency (\%) 
  \\ \hline \hline 
0.1000E-03 & 0.1778E-03 & 0.439E+01  & 0.106E+01 & 24.1\\ \hline
0.1778E-03 & 0.3162E-03 & 0.414E+01  & 0.113E+01 & 27.3\\ \hline
0.3162E-03 & 0.5623E-03 & 0.360E+01  & 0.961E+00 & 26.7\\ \hline
0.5623E-03 & 0.1000E-02 & 0.296E+01  & 0.704E+00 & 23.8\\ \hline
0.1000E-02 & 0.1778E-02 & 0.232E+01  & 0.396E+00 & 17.1\\ \hline
0.1778E-02 & 0.3162E-02 & 0.174E+01  & 0.171E+00 & 9.8\\ \hline 
0.3162E-02 & 0.5623E-02 & 0.125E+01  & 0.381E-01 & 3.0\\ \hline 
0.5623E-02 & 0.1000E-01 & 0.853E+00  & 0.189E-02 & 0.22\\ \hline \hline
\end{tabular}
\caption{\it \label{tab:lowq2}
Cross sections with and without the cuts mentioned in
  Eq. (\ref{eq:cuts}) for $2 < Q^2 < 10$ GeV$^2$.} 
\end{table}
In the low $x$ range, we observe an efficiency of 20 -- 25 \%,
diminishing near threshold, the low efficiency mostly a result of
the $p_t$ cut. 

Table~\ref{tab:medq2} displays results in the medium $Q^2$ range, from
10 to 100 GeV$^2$.
\begin{table}
% 10 - 100 GeV^2
\begin{tabular}{||c|c||c|c|c||} \hline \hline
  $x_{min}$ & $x_{max}$ &  $\sigma$ (nb) & $\sigma_{cuts}$ (nb) &
  Efficiency (\%) 
  \\ \hline \hline 
0.1000E-03 & 0.1778E-03 &  0.334E+00 & 0.584E-01 & 17.5\\ \hline 
0.1778E-03 & 0.3162E-03 &  0.116E+01 & 0.251E+00 & 21.6\\ \hline 
0.3162E-03 & 0.5623E-03 &  0.173E+01 & 0.484E+00 & 28.0\\ \hline 
0.5623E-03 & 0.1000E-02 &  0.198E+01 & 0.681E+00 & 34.4\\ \hline 
0.1000E-02 & 0.1778E-02 &  0.191E+01 & 0.768E+00 & 40.2\\ \hline 
0.1778E-02 & 0.3162E-02 &  0.162E+01 & 0.701E+00 & 43.3\\ \hline 
0.3162E-02 & 0.5623E-02 &  0.127E+01 & 0.504E+00 & 39.7\\ \hline
0.5623E-02 & 0.1000E-01 &  0.945E+00 & 0.280E+00 & 29.6\\ \hline 
0.1000E-01 & 0.1778E-01 &  0.660E+00 & 0.112E+00 & 17.0\\ \hline 
0.1778E-01 & 0.3162E-01 &  0.430E+00 & 0.254E-01 & 5.9\\ \hline 
0.3162E-01 & 0.5623E-01 &  0.256E+00 & 0.136E-02 & 5.3\\ \hline 
0.5623E-01 & 0.1000E+00 &  0.131E+00 & 0.468E-05 & 0.0036\\ \hline \hline
\end{tabular}
\caption{\it \label{tab:medq2}
Cross sections with and without the cuts mentioned in
  Eq. (\ref{eq:cuts}) for $10 < Q^2 < 100$ GeV$^2$.}
\end{table}
We find the efficiency has risen considerably to the 30--40 \%
range where the cross section is peaked in intermediate $x$, with on
the order of 20 \% for the small $x$ region.

In the large $Q^2$ range, from 100 to 1000 GeV$^2$, the results are
found in Tab.~\ref{tab:highq2}.
\begin{table}
% 100 - 1000 GeV^2
\begin{tabular}{||c|c||c|c|c||} \hline \hline
  $x_{min}$ & $x_{max}$ &  $\sigma$ (nb) & $\sigma_{cuts}$ (nb) &
  Efficiency (\%) 
  \\ \hline \hline 
0.1000E-02 & 0.1778E-02 &  0.401E-01 & 0.120E-01 & 29.9\\ \hline 
0.1778E-02 & 0.3162E-02 &  0.124E+00 & 0.486E-01 & 39.2\\ \hline 
0.3162E-02 & 0.5623E-02 &  0.159E+00 & 0.885E-01 & 55.7\\ \hline 
0.5623E-02 & 0.1000E-01 &  0.157E+00 & 0.958E-01 & 61.0\\ \hline 
0.1000E-01 & 0.1778E-01 &  0.128E+00 & 0.757E-01 & 59.1\\ \hline  
0.1778E-01 & 0.3162E-01 &  0.906E-01 & 0.458E-01 & 50.6\\ \hline 
0.3162E-01 & 0.5623E-01 &  0.572E-01 & 0.173E-01 & 30.2\\ \hline 
0.5623E-01 & 0.1000E+00 &  0.315E-01 & 0.285E-02 & 9.1\\ \hline 
0.1000E+00 & 0.1778E+00 &  0.140E-01 & 0.167E-03 & 1.2\\ \hline 
0.1778E+00 & 0.3162E+00 &  0.407E-02 & 0.400E-05 & 0.098\\ \hline 
0.3162E+00 & 0.5623E+00 &  0.466E-03 & 0.413E-08 & 0.0009\\ \hline \hline
\end{tabular}
\caption{\it \label{tab:highq2}
Cross sections with and without the cuts mentioned in
  Eq. (\ref{eq:cuts}) for $100 < Q^2 < 1000$ GeV$^2$.}
\end{table}
The efficiency continues to rise with $Q^2$, reaching 50--60 \% in the
high to intermediate $x$ region and 30--40 \% in the small $x$ range.

The number of events landing within the imposed cuts mentioned in
Eq. (\ref{eq:cuts}) is not overwhelmingly large, but a large enough
sample should be accessible to gather significant statistics.

\subsection{Charm Production at Asymptotic $Q^2$}

Near the threshold for charm production the deep inelastic
structure functions $F_i$, $i=2,L$, which include the contributions of the 
light partons $u, \, d, \, s,$ and $g$ and the charm quark with mass
$m_c$ are given by
%(2.1)
\begin{eqnarray}
\label{jack1}
&&F_i(Q^2,m_c^2,3) =
\frac{2}{9}\Big[\Sigma(\mu^2,3)
\otimes \Big\{ C^{\rm S}_{i,q}\Big(\frac{Q^2}{\mu^2} , 3\Big)
+ L^{\rm S}_{i,q}\Big(\frac{Q^2}{m_c^2}, \frac{m_c^2}{\mu^2},3\Big)\Big\} \Big]
\nonumber \\ && \qquad \qquad\qquad
+ G(\mu^2,3) \otimes\Big\{ C^{\rm S}_{i,g}\Big(\frac{Q^2}{\mu^2}, 3\Big)
+ L^{\rm S}_{i,g}\Big(\frac{Q^2}{m_c^2}, \frac{m_c^2}{\mu^2},3\Big)\Big\}
\nonumber \\ && \qquad \qquad\qquad
+ \Delta(\mu^2,3) \otimes\Big\{ C^{\rm NS}_{i,q}\Big(\frac{Q^2}{\mu^2}, 3\Big)
+ L^{\rm NS}_{i,q}\Big(\frac{Q^2}{m_c^2}, \frac{m_c^2}{\mu^2},3\Big)\Big\}
\nonumber \\ && \qquad \qquad\qquad
+ \frac{4}{9}\Big[\Sigma(\mu^2,3)
\otimes H^{\rm PS}_{i,q}\Big(\frac{Q^2}{m_c^2} ,\frac{m_c^2}{\mu^2}, 3\Big)
\nonumber \\ && \qquad \qquad\qquad
+ G(\mu^2,3)  \otimes H^{\rm S}_{i,g}
\Big(\frac{Q^2}{m_c^2}, \frac{m_c^2}{\mu^2},3\Big) \Big] \,,
\end{eqnarray}
where we have the following definitions. The variable $Q^2$
denotes the virtual mass squared of the photon exchanged between 
the electron and the proton.. The momenta of the photon and the 
proton are given by $q$ and $p$ respectively
and the Bjorken scaling variable is defined by $x = Q^2/(2p \cdot q)$
$ (q^2 =  -Q^2 <0)$.
The structure functions $F_i$ depend on $x$, on $Q^2$ and on the 
charm quark mass $m_c$.
The convolution symbol $\otimes$ appears on the right-hand-side 
which is defined by
%(2.2)
\begin{eqnarray}
\label{jack2}
(f\otimes g) (x) = \int_0^1 \, dz_1 \int_0^1 \, dz_2 \,\delta(x - z_1 z_2)
\,f(z_1)\, g(z_2) \,.
\end{eqnarray}
The gluon density is denoted by $G(\mu^2,3)$ where
the number of light flavours $n_f$ is taken to be three.
Also the singlet and nonsinglet combinations of the light quark
densities are defined for $n_f = 3$ as follows
%(2.3)
\begin{eqnarray}
\label{jack3}
&&\Sigma(\mu^2,3) = u(\mu^2,3)+ \bar u(\mu^2,3)
+ d(\mu^2,3) + \bar d(\mu^2,3)
\nonumber \\ && \qquad \qquad\qquad
+ s(\mu^2,3) + \bar s(\mu^2,3) \,,
\end{eqnarray}
%(2.4)
\begin{eqnarray}
\label{jack4}
&&\Delta(\mu^2,3) = \frac{2}{9} \Big[u(\mu^2,3)+ \bar u(\mu^2,3)\Big]
 - \frac{1}{9} \Big[ d(\mu^2,3) + \bar d(\mu^2,3) 
\nonumber \\ && \qquad \qquad
+ s(\mu^2,3) + \bar s(\mu^2,3)\Big] \,.
%\nonumber \\ && 
\end{eqnarray}
The same singlet and non-singlet classification can also be made
for the light parton coefficient functions 
$C_{i,k}$ ($i=2,L, k= q,g$) and the heavy quark coefficient
functions $L_{i,k}$, $H_{i,k}$, $H_{i,k}$ depending on $m_c$. The
coefficient functions $L_{i,k}$, $H_{i,k}$  
reflect the corresponding production processes.
The functions $H_{i,k}$ contain the processes where 
the virtual photon couples to the heavy quark 
({\it c.f.} Figs. 2, 3, 4, 5.a and 5.b in \cite{lrsn1}), whereas
$L_{i,k}$ describe the reactions where the
virtual photon couples to the light quark ({\it c.f.} Figs. 5.c and 5.d
in \cite{lrsn1}).  
Hence $L_{i,k}$ and $H_{i,k}$ are multiplied by
$e_i^2$ (where $1/n_f \sum _{i=1}^3 e_i^2 = 2/9$)
and by $e_c^2$ (where $e_c = 2/3)$ respectively.
These charge factors are exhibited by Eq. (\ref{jack1}).
Furthermore both the coefficient functions and the parton densities depend on 
the mass factorization scale $\mu\,$, which, for convenience will be set 
equal to the renormalization scale. The latter 
also shows up in the running coupling
constant $\alpha_s(\mu^2,3)$ where the number of light flavours is three.

Equation (\ref{jack1}) gives an adequate prescription as long as the
c.m. energy 
is not too far above the charm threshold, which implies that $Q^2$ is
not too large compared to $m_c^2$. However, when we enter the asymptotic
region $Q^2 \gg m_c^2\,$, the heavy quark coefficient functions
behave like $\ln^i (m_c^2/\mu^2) \ln^j (Q^2/m_c^2)$ so that the
higher order corrections can become large.
%This implies
%a problem with the convergence of the perturbation
%series, which has to be remedied. 
At sufficiently large $Q^2$ the
charm quark should be treated in the same way as the light partons
were at smaller $Q^2$.
The logarithmic behaviour of the coefficient functions is due to the
collinear singularities which are regulated by $m_c$.
Therefore when $Q^2 \gg m_c^2$, the charm quark behaves like a 
massless quark similar to the behaviour of the normal light 
quarks ($u$, $d$, and $s$) over the whole $Q^2$ range.
Following the same procedure as has been used for the light partons,
the mass singular $(m_c$ -dependent) terms have to be factorized
out of the heavy quark coefficient functions using the method of 
mass factorization.
This leads to a redefinition of the parton densities in Eqs. (\ref{jack3}),
(\ref{jack4}) 
and the heavy quark coefficient functions turn into the light parton
analogues, wherein the number of light flavours is enhanced by one.
The above procedure is called the variable flavour number scheme (VFNS) 
which is outlined in leading order in \cite{aot}.

Since all coefficient functions are now available
up to order $\alpha_s^2$, this analysis can be extended to NLO to give
a better description for the structure functions $F_i(x,Q^2)$ at large
$Q^2$.  A preprint is in preparation\cite{buza3}.

\subsection{Bottom Quark Production}

\begin{figure}[tb]
\centering 
\epsfig{file=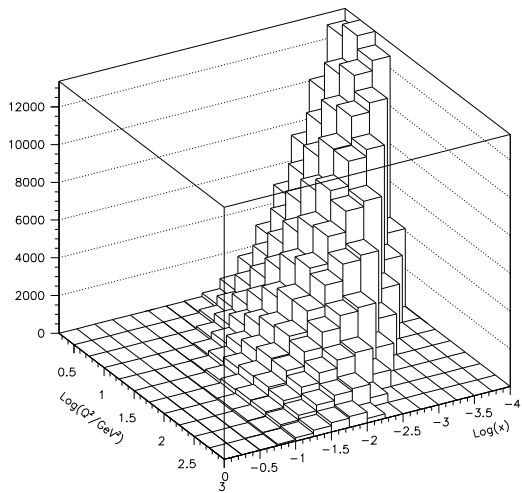,width=7.5cm,angle=0}
\caption{\label{fig:bevents}
{\it
Projected number of DIS bottom events for an integrated luminosity of
500 $pb^{-1}$ binned in $x$ and $Q^2$ with no cuts applied.
  }}
\end{figure}

The anticipated amount of bottom production is greatly reduced due to
the reduction of the charge factor by four as well as a significantly
reduced phase space.  Given a luminosity of 500 $pb^{-1}$, however,
thousands of events are expected per bin of $x$ and $Q^2$ as
shown in Fig. \ref{fig:bevents}.  We take $m_b = 4.75$ GeV, the
factorization scale $\mu^2 = Q^2 + m_b^2$, as well as the CTEQ3M
distributions. 
A discussion of the implications of this number of $b$ events can be
found in section \ref{sec:expbottom}. 

\section{Experimental Aspects}
\label{sec:ea}
\subsection{Current Experimental Situation}
Results on charm production in deep inelastic $ep$ scattering are available 
from both experiments H1 and ZEUS based on an integrated 
luminosity of roughly 3~$pb^{-1}$ collected with each experiment
at HERA in 1994. 
The H1 collaboration \cite{H1} has performed the tagging of heavy quark 
events by reconstructing $D^0$(1864)\footnote{Charge conjugate states are 
always included.} and $D^{*+}$(2010) mesons,
while the ZEUS collaboration \cite{ZEUS} has given preliminary results 
for the inclusive $D^{*+}$(2010) analysis.
   
The $D^0$ is identified via its decay mode
$D^0\rightarrow K^-\pi^+$
and the $D^{*+}$ through the decay chain
$D^{*+}\rightarrow D^0\pi^+_{slow}\rightarrow K^-\pi^+\pi^+_{slow}$.
For the latter use is made of the tight kinematic constraint for the decay of
$D^{*+}\rightarrow D^0\pi^+_{slow}$ \cite{feldmann}. 
A better resolution is expected for the mass difference
$ \Delta m = m(D^0 \pi^+_{slow}) - m(D^0)\label{deltam}$
than for the $D^{*+}$ mass itself.

\begin{figure}[t] 
\begin{picture}(15.,4.8)
\put(0,-0.8){\epsfig{file=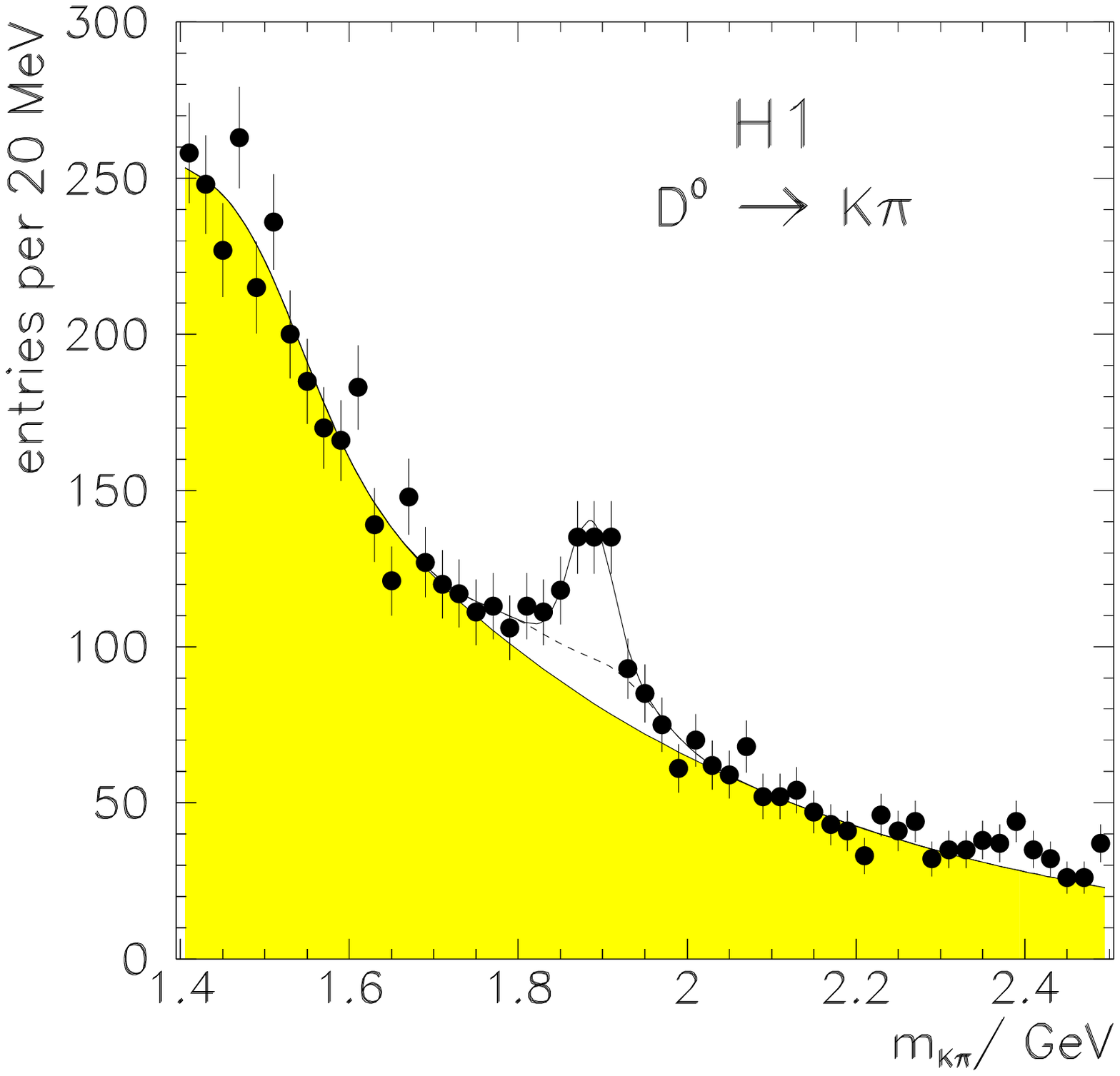,width=6.5cm}}
\put(4.,-8.25){\epsfig{file=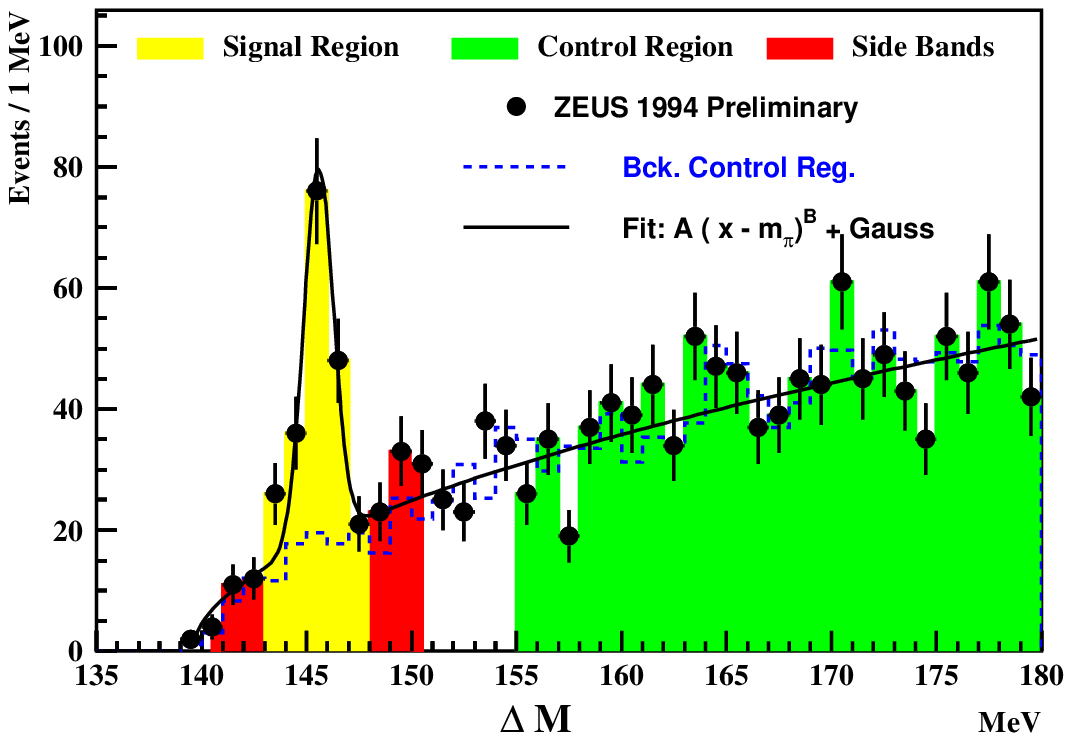,width=14.6cm,clip=}}
\put(2.,3.8){(a)}
\put(8.5,3.8){(b)}
\end{picture}
\normalsize
\caption[junk]{\label{fig1}{\it
{\rm (a)} The $K^-\pi^+$ mass distribution observed in the inclusive $D^0$
analysis of DIS events from H1, and {\rm (b)} the $\Delta m$ distribution
         obtained in the inclusive $D^{*+}$ analysis  of DIS events
         from ZEUS.  
  }}
\end{figure}

Figure \ref{fig1} shows the $m_{K\pi}$ distribution obtained in the inclusive 
$D^0$ analysis of H1 and the $ \Delta m$ distribution as 
observed in the inclusive $D^{*+}$ of ZEUS. Evidently the number of observed 
events containing heavy quarks is small. Only of the order of 100 to 200 
charm mesons are identified in any of the different analyses. Combining the 
$D^0$ and $D^{*\pm}$ analysis of H1 leads to a charm production
cross section of
\begin{equation}
\sigma(ep\rightarrow ec\overline c X)=17.4\pm1.6\pm1.7\pm1.4~\mbox{nb}
\end{equation}
in the kinematic range $10\;\gev^2<Q^2<100\;\gev^2$ and $0.01<y<0.7$. 
The errors refer to the statistical, the experimental systematic and the model
dependent error, respectively. The model dependent error accounts for the 
uncertainty in the determination of acceptances due to the choice of the parton
density in the proton, the mass of the charm quark, and the fragmentation 
function.
This cross section is somewhat larger than predicted by the NLO calculations 
\cite{lrsn1,H1}. The prediction from the gluon density
in the proton  
extracted from the NLO QCD fit of H1~\cite{h1fit} comes closest to the 
charm data.   
\unitlength1cm

\begin{figure}[tb] 
\begin{picture}(15.,4.8)
\put(5,-.8){\epsfig{file=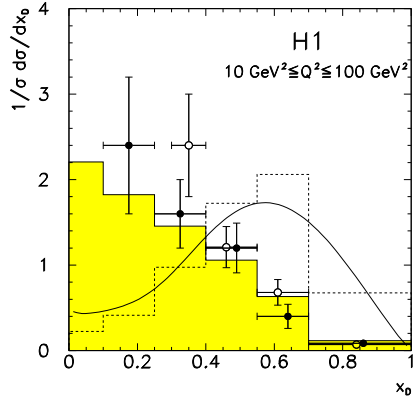,width=6.5cm}}
\end{picture}
\normalsize
\caption[junk]{\label{fig2}{\it
 Normalized $x_D$ distribution in deep inelastic 
          $ep$ scattering 
          at $\langle W\rangle\approx 125\;\gev$ for $|\eta_D|<1.5$.
          The open/closed points represent the $D^0$/$D^*$ data 
          of H1.          
          The shaded histogram shows the BGF expectation 
          according to AROMA. 
          The dashed histogram
          shows the charm sea expectation, obtained by
          selecting QPM events from LEPTO. 
          The full line gives the result of the
          QCD evolution of the 
          fit to the charged current 
          $\stackrel{\scriptscriptstyle(-)}{\textstyle \nu}N$ data.
  }}
\end{figure}
Information on the charm production mechanism in neutral current DIS at HERA
is obtained from the distribution of $x_D=2|\vec P^*_{D^0}|/W$,
where $\vec P^*_{D^0}$ denotes the momentum of the 
$D^0$ in the $\gamma^*p$ system. 
Figure~\ref{fig2} shows the distributions ${1}/{\sigma}\; {d \sigma}/{dx_D}$, 
which are a convolution of the charm production spectrum with the
fragmentation function compared to the LO BGF expectation of the AROMA 
generator. 
%Good agreement is observed between the $x_D$
%distribution obtained from  the two different H1 analyses. 
The BGF model %, where 
%two charm quarks are recoiling against the proton remnant
%in the hadronic cms, 
 agrees very well with the shape of the data.
The figure also includes the expectations for charm mesons originating from
quarks in the proton either by using 
LEPTO 6.1 generator~\cite{lepto}, from which only charm sea quark events 
are selected, or by extrapolating the results from charm production in
charged current $\stackrel{\scriptscriptstyle(-)}{\textstyle \nu}N$ scattering 
\cite{neutrino} to HERA energies.
Large differences in the shape of the distributions are observed for the data 
and these QPM expectations. From these differences, it is concluded
that more than 95\% 
of the charm production in neutral current deep inelastic $ep$ scattering 
is due to boson-gluon fusion.
This observation seems to be in contradiction to recent inclusive calculations
of charm production in DIS \cite{aot}, from which was concluded that for 
the kinematic range accessible in the current analyses at HERA charm quarks 
may already be considered as partons in the proton \cite{lbhmmrsv}. 
   
\begin{figure}[t]
\begin{picture}(17.8,5.3)
\put(3.0,-3.1){\epsfig{file=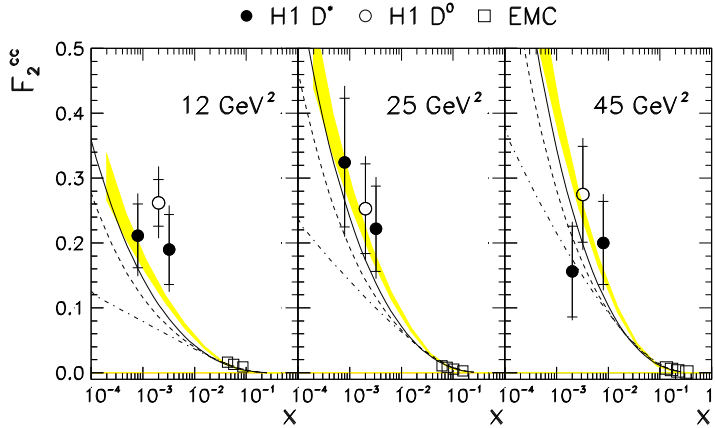,width=11cm}}
\end{picture}
\centering
\caption{ {\it  $F_2^{c\overline c}$ as derived from the inclusive
          $D^{*+}$ (full dots) and $D^0$ analysis (open circles)
 in comparison
          with the NLO calculations based on GRV-HO (full line), 
          MRSH (dashed line), and MRSD0$^\prime$ (dash-dotted line)
          parton distributions
          using a charm quark mass of $m_c=1.5\;\gev$ for 
          $\langle Q^2\rangle=12,\;25\;{\rm and }~45\;\gev^2$. 
          The inner (outer) error bars refer to the statistical 
          (total) errors. The shaded band represents the prediction from the 
          H1 NLO fit to the $F_2$ measurements.
          The EMC data are also shown (open boxes).
          }}
\label{fig4}
\end{figure}

The measurements of the charm contribution 
$F^{c\overline c}_2(x,Q^2)$ to the proton structure function derived from the
inclusive $D^0$ and $D^{*+}$ analysis of H1 are displayed in Fig.~\ref{fig4}
together with the results of the EMC collaboration \cite{emc} at large $x$.
The measurement at HERA extends the range of the $F_2^{c\overline c}$ 
measurement by two orders of magnitude towards smaller $x$ values. The 
comparison of the H1 and EMC measurements reveals a steep rise of 
$F_2^{c\overline c}$ with decreasing $x$. The data are compared with  
NLO calculations \cite{lrsn1} using the GRV-HO
\cite{grv92}, the 
MRSH \cite{mrsh}, and the  
MRSD0$^\prime$ \cite{mrsd} parameterizations of the gluon density in
the proton for a  
charm quark mass $m_c=1.5\;\gev$.
The data are also compared to the prediction from the H1 QCD fit to the 
$F_2$ measurements using a charm quark mass of $m_c=1.5\;\gev$. The error
band shown for this fit includes the propagation of the statistical and the
uncorrelated systematic errors on the total $F_2$ data through the fitting
procedure. This prediction lies systematically above all other calculations, 
independently of $x$ and $Q^2$ but agrees better with the 
$F_2^{c\overline c}$ measurement. Within the errors the 
ratio of $\left\langle F_2^{c\overline c}/F_2\right\rangle$ is found to be 
independent of $x$ and $Q^2$. Averaged over the kinematic range a ratio
$\left\langle F_2^{c\overline c}/F_2\right\rangle=0.237\pm0.021\pm0.041$ is 
obtained, which is one order of magnitude larger than at larger $x$ measured 
by EMC.

\subsection{Future}
\subsubsection{Charm Production}
In this section some estimates about the precision that may be expected at 
HERA for an integrated luminosity  of 500~$pb^{-1}$ are given.
As an example the capabilities of the H1 detector \cite{detector} are 
considered. For heavy flavor physics an important new feature of the apparatus
is the double layer silicon vertex detector (CST)~\cite{cst} which will 
allow use of the apparent proper time of charm hadrons in selecting
heavy flavor events.. Other decay channels than $D^0\rightarrow K^-\pi^+$ and 
$D^{*+}\rightarrow D^0\pi^+_{slow}\rightarrow K^-\pi^+\pi^+_{slow}$ are already
under investigation or will become accessible due to the CST.

Table~\ref{tab1} contains a list of decay modes which should be feasible for 
charm tagging in H1 in future by also including the CST. A total charm 
selection efficiency of 1\% may be obtained. Compared to the 
present analysis this corresponds to an increase in the total selection 
efficiency of a factor 4 to 5. Although the new channels opened by
the use of the CST corresponds only to 50\% of this gain, the signal to 
background ratio will improve for all decay modes considerably. 
Because of this the effective 
number of events per luminosity $N_{eff}=(N_{signal}/\sigma_{stat})^2$ will 
increase by a factor of 6 to 8.
Due to the cut
in the impact parameter of the $D^0$ mesons, which is likely to be different in
the $D^0$ and $D^{*+}$ analyses, the events selected in the $D^{*+}$ analyses
will only partially overlap with the events selected in the
corresponding  $D^0$ analyses.
Taking an integrated luminosity of 500 pb$^{-1}$ about 160,000 tagged charm
events are expected for the kinematic range of the published H1 
analysis. This has to be compared to the currently analyzed 250 events.
Taking into account the increase in the effective number of events a gain of a
factor of 1000 is expected for the statistical significance.  
In total we expect to observe roughly 6,000 double tag charm events in the
range $1.7~\mbox{GeV}^2<Q^2<560~\mbox{GeV}^2$ which would allow a
study of the 
charm production dynamics in detail (see Ref.~\cite{lbhmmrsv}).
\begin{table}[t]
\centering
\begin{tabular}{|l|c|r|r|r|r|}
\hline
&&&&&\cr
Mode&&$P(c\rightarrow D)$&$BR(D\rightarrow FS)$&$\epsilon_{tot}$&
$P\cdot BR\cdot\epsilon_{tot}$\cr 
&&&&&\cr
\hline\hline
%\noalign{\smallskip}
$D^{*+}\rightarrow D^0\pi^+_{s}\rightarrow K^-\pi^+\pi^+_{s}$
&\cite{H1}&0.248&0.026&0.16&0.0010\cr
$D^{*+}\rightarrow D^0\pi^+_{s}\rightarrow K^-\pi^+\pi^0\pi^+_{s}$
&&&0.096&0.04&0.0010\cr
$D^{*+}\rightarrow D^0\pi^+_{s}\rightarrow K^-3\pi\pi^+_{s}$
&&&0.052&0.10&0.0013\cr
$D^{*+}\rightarrow D^0\pi^+_{s}\rightarrow K^0\pi^+\pi^-\pi^+_{s}$
&&&0.013&0.20&0.0007\cr
$D^{*+}\rightarrow D^0\pi^+_{s}\rightarrow K^0\pi^+\pi^-\pi^0\pi^+_{s}$
&&&0.024&0.08&0.0005\cr
$D^{*+}\rightarrow D^0\pi^+_{s}\rightarrow K^-\mu^+\nu_\mu\pi^+_{s}$
&&&0.024&0.04&0.0002\cr
$D^{*+}\rightarrow D^0\pi^+_{s}\rightarrow K^-e^+\nu_e\pi^+_{s}$
&&&0.024&0.04&0.0002\cr
%\noalign{\smallskip}
\hline
%\noalign{\smallskip}
Sum $D^{*+}$&&0.248&0.259&&0.0049\cr
%\noalign{\smallskip}
\hline\hline
%\noalign{\smallskip}
$D^0\rightarrow K^-\pi^+$&\cite{H1}&0.535&0.0383&0.06&0.0012\cr
$D^0\rightarrow K^0\pi^+\pi^-$&&&0.0186&0.06&0.0006\cr
$D^0\rightarrow K^-3\pi$&CST&&0.075&0.04&0.0014\cr
%\noalign{\smallskip}
\hline
%\noalign{\smallskip}
Sum $D^0$ &&0.535&0.132&&0.0032\cr
%\noalign{\smallskip}
\hline\hline
$D^+\rightarrow K^0\pi^+$&CST&0.25&0.009&0.12&0.0003\cr
$D^+\rightarrow K^-\pi^+\pi^-$&CST&&0.091&0.07&0.0016\cr
$D^+\rightarrow K^03\pi$&CST&&0.07&0.07&0.0012\cr
\hline
Sum $D^+$&CST&0.25&0.170&&0.0031\cr
\hline
\hline
Sum $D$&&&&&0.0097\cr
\hline

%\noalign{\smallskip}
\end{tabular}
\caption{\label{tab1} {\it Compilation of various decay channels of charm 
mesons accessible in H1. The channels marked by ``CST" will be
accessible only by using the silicon vertex detector. $BR(D\rightarrow FS)$ 
denotes the product of all branching ratios involved in the decay into the 
final state ``FS". For the determination of the total charm tagging efficiency
the correlations in the $D^0$ and $D^{*+}$ analyses are taken into account.
}}
\end{table} 
\begin{figure}[t]
\begin{picture}(17.8,6.3)
\put(1.0,-1.7){\epsfig{file=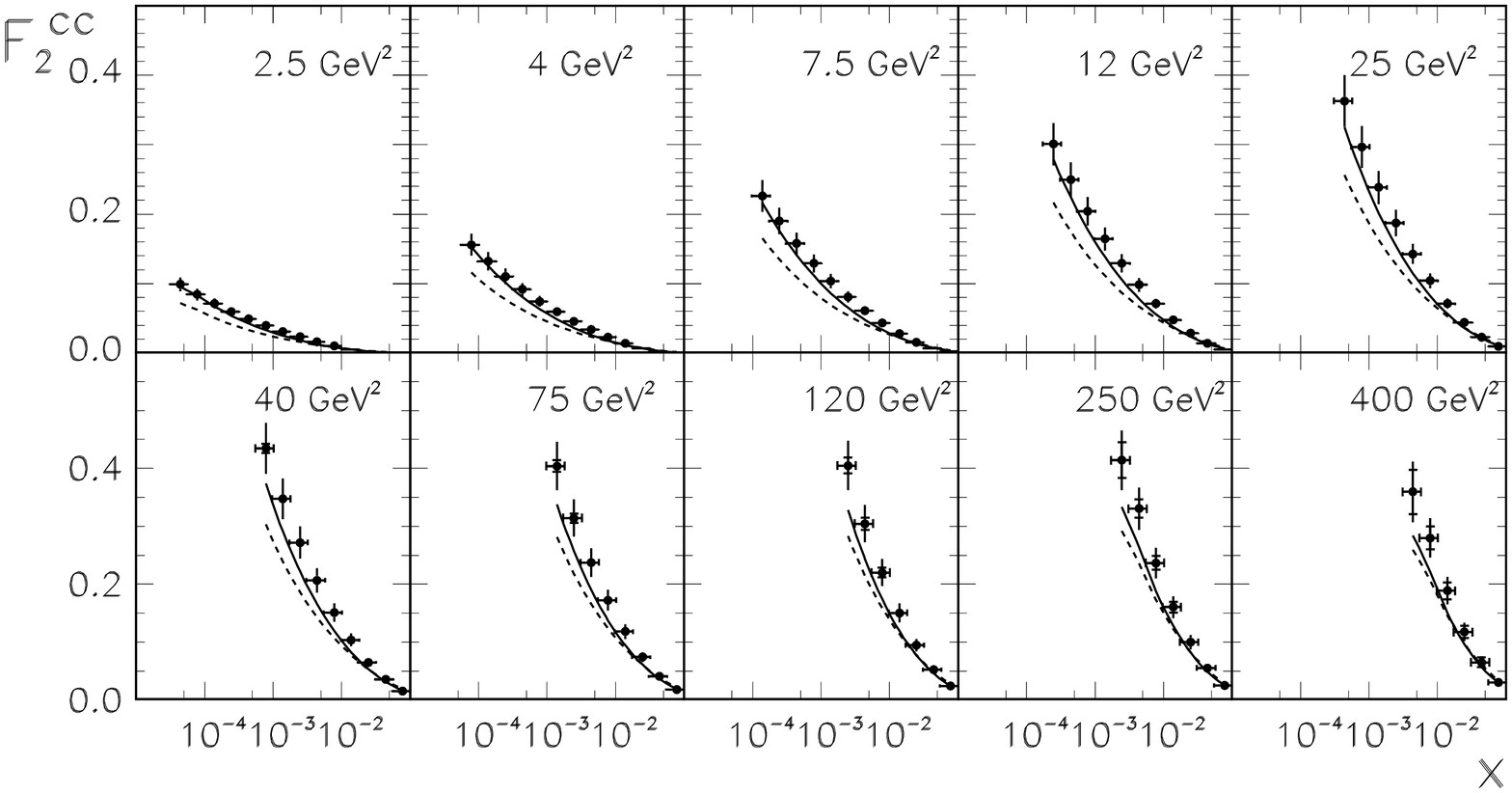,width=15cm}}
\end{picture}
\centering
\caption{ {\it Expected $F_2^{c\overline c}$ for a luminosity of 500
$pb^{-1}$. 
          The points show the prediction from the gluon density determination
          by the NLO H1 fit to the inclusive $F_2$. The inner (outer) error 
          represents the statistical and the total experimental error. 
          The full (dashed) line gives the
          expectation from the NLO calculations based on GRV-HO and 
          MRSH parton distributions
          using a charm quark mass of $m_c=1.5\;\gev$.
          }}
\label{figf2cc}
\end{figure}

Figure~\ref{figf2cc} shows the result of the hypothetical measurement of 
$F_2^{c\overline c}$ in the range $1.7~\mbox{GeV}^2<Q^2<560~\mbox{GeV}^2$
for a luminosity of 500 pb$^{-1}$ based on the gluon 
density determination from the NLO H1 fit to the inclusive $F_2$ data combining
the statistics of the $D$ meson decay modes summarized in
Table~\ref{tab1} by assuming $m_c=1.5~\mbox{GeV}$ and
$\mu^2=Q^2+4m_c^2$. 
The contribution due to
$F_L$ and the light flavors are not included.
Statistical and full errors, obtained by adding the statistical and 
experimental systematic error in quadrature are shown. The change in the 
acceptance due to the value of $m_c$ and $\mu$ is not included in the error.
For $Q^2<100~\mbox{GeV}^2$ the precision of this measurement will be limited 
by the experimental systematic uncertainty.   

A detailed analysis of the experimental systematic uncertainties in the 
determination of $F^{c\overline c}_2$ with the current experimental 
statistics can be found in Ref. \cite{H1} for the H1 analysis. At present
the total systematic error is approxiamtely 20~\%. It is dominated by the 
uncertainty in the assumptions made to extract the signal from the observed 
mass distributions. 
This is clearly an effect of the limited statistics available in the current 
analyses.
Because of the increase in statistical significance of the data with a
luminosity of 500~$pb^{-1}$ a much better understanding of 
the measured mass distributions as well as of the detector is expected.
Ultimately a systematic error of 10\% is achievable, which will then be to 
equal parts due to detector and analysis related errors (7\%)
and to the knowledge of the fragmentation probability $P(c\rightarrow D)$ 
of a charm quark into a specific charm meson and their branching ratios
(7\%). For the latter the experimental situation is not expected to improve 
in the near future. 

Figure~\ref{figf2cc} also shows the NLO predictions using the GRV-HO(1992) and 
the MRSH
parameterization of the gluon density in the proton using 
$m_c=1.5~\mbox{GeV}$ and $\mu^2=Q^2+4m_c^2$. For a given value of $m_c$
the data will still allow a 
sensitive indirect determination of the gluon density even with the 
relatively large experimental systematic uncertainties.
Due to the high statistics it will be possible to measure the charm production
cross section up to $Q^2\ge1000~\mbox{GeV}^2$. Assuming that charm tagging is 
performed for all decay modes listed in Tab.~\ref{tab1}, the sensitivity limit
for the inclusive measurements will already be reached at 50~$pb^{-1}$ for 
$Q^2\le100~\mbox{GeV}^2$, at which time the experimental systematic
will dominate in this kinematic range. 

\begin{figure}[t]
\begin{picture}(15,4.8)
\put(4.0,-0.8){\epsfig{file=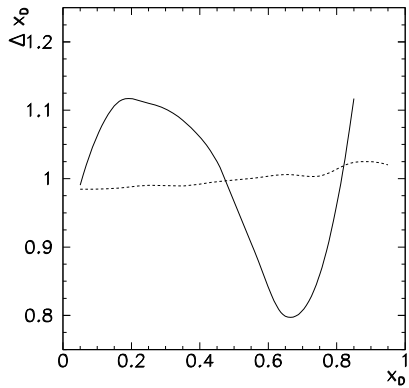,width=6.5cm}}
\end{picture}
\centering
\caption{ {\it 
          Relative change $\Delta x_D$ in the shape of the distribution  
          ${1}/{\sigma}\; {d \sigma}/{dx_D}$ according to the LO Monte Carlo
          Simulation of the AROMA program for 
          $6~\mbox{GeV}^2<Q^2<100~\mbox{GeV}^2$ and the cuts of 
          Eq.~(\ref{eq:cuts}).
          The full line gives the change in the shape by changing $m_c$ 
          from $1.3~\mbox{GeV}$ to $1.7~\mbox{GeV}$. The dashed line
          shows the influence of using the GRV-HO (1992) instead of the MRSH
          parameterization of the gluon density in the proton.
          }}
\label{fig5}
\end{figure}

The discussion in the theory section has shown, that the measurement of 
$F_2^{c\overline c}$ is sensitive to the gluon density to some extent
at large $y$.  
Unfortunately the inclusive measurement in this range is also very sensitive 
on the charm quark mass. Therefore it is questionable whether the 
measurement of $F_2^{c\overline c}$ alone will allow an extraction of the 
gluon density at small $x$. In the experimental analysis of
Ref. \cite{H1} it was  
observed that the change in $F_2^{c\overline c}$ at small $Q^2$ due to $m_c$
is compensated by the change in the acceptance such that the
measurement should still allow the extraction of the gluon density
from the inclusive measurement. The behavior of the acceptance at
$Q^2<10~\mbox{GeV}^2$ has not been
studied yet. If it happens that $F_2^{c\overline c}$ does not provide an
reliable indirect extraction of the gluon density because of the uncertainty
in $m_c$, exclusive distributions of the identified charm hadrons have to be
studied. As an example the result of the Monte Carlo study of the influence 
of $m_c$ and the gluon density on the distribution 
${1}/{\sigma}\; {d \sigma}/{dx_D}$ is shown in Fig. \ref{fig5} for 
$6~\mbox{GeV}^2<Q^2<100~\mbox{GeV}^2$ and the cuts of Eq.~(\ref{eq:cuts}).
Here $\Delta~x_D$ is defined as
\begin{equation}
\Delta~x_D=\frac{{1}/{\sigma_i}\; {d \sigma_i}/{dx_D}(x_D)}
{{1}/{\sigma_j}\; {d \sigma_j}/{dx_D}(x_D)}
\end{equation}
where $i,j$ denote the different values of the parameters, i.e. 
$m_c=1.3,~1.7~\mbox{GeV}$ and PDF=MRSH, GRV-HO(1992), respectively.
The mass of the charm quark affects strongly the shape of the 
$x_D$~distribution (full line) while there is only very little effect due
to the choice of the parton density function (dashed line). 
A study of this distribution, for instance, should therefore 
disentangle the effect of the charm quark mass and the gluon 
density on $F_2^{c\overline c}$.

In section 2.5, the question when charm quarks should be treated as
partons of the proton was discussed.
The investigation of the $x$ and $Q^2$ dependence of the 
exclusive measurement of ${1}/{\sigma}\; {d \sigma}/{dx_D}$
(see Fig.~\ref{fig2}) will allow a study of the 
origin of charm production as a function of the kinematic variables. At 
sufficiently large $Q^2$ it is expected that 
the charm quark behaves like a parton in the proton. This will result in
a change of the $x_D$ distribution from the boson gluon fusion dominated 
regime presently observed in the available data at
$\langle Q^2\rangle\approx25~\mbox{GeV}^2$ to the 
sea quark dominated regime at large $Q^2$.
If it becomes possible to control the $\mu$ and $m_c$ dependence of
charm production at large $Q^2$, the measurement of 
$F_2^{c\overline c}(x,Q^2,x_D)$ can determine
$g(x,Q^2)$ and $c(x,Q^2)$ simultaneously. This study would certainly require 
a luminosity of 500 $pb^{-1}$ to produce enough data at large $Q^2$. 

\subsubsection{Bottom Production}
\label{sec:expbottom}
In section 2.6, predictions for the number of bottom events as a
function of $x$ 
and $Q^2$ were given for a luminosity of 500~pb$^{-1}$. 
In the following the possibilities to measure $F_2^{b\overline b}(x,Q^2)$ 
at HERA will be discussed using again the H1 detector as an example. 

Compared to charm quark events
the major experimental difference in bottom quark production 
is the relatively long lifetime of the $B$ mesons. With use of the
CST, bottom events are selected by applying a cut in the impact parameter of
tracks not fitting to the primary event vertex. The combinatorial background
is negligible for this analysis, while the charm production is a significant
background source, because of the much larger cross section.
 Two different possibilities will be discussed here, namely
\begin{enumerate}
\item the exclusive analysis of reconstructed $D$ mesons. This method
benefits from
\begin{enumerate}
\item the branching of $B$ mesons into charm meson being very large
\cite{pdf}, i.e. 
$BR(B\rightarrow D^\pm X)=0.242\pm0.033$, 
$BR(B\rightarrow D^0/\overline D^0X)=0.58\pm0.05$ and
$BR(B\rightarrow D^{*\pm})X=0.0231\pm0.033$, and
\item the relatively long visible lifetime observed in
these decay chains. Taking into account
the decay length of both the $B$ mesons and the subsequent $D$ mesons,
values of $c\tau\approx 590~\mu m$ and $c\tau\approx 780~\mu m$ are obtained
for the decay chains $B\rightarrow D^0/\overline D^0X$ and 
$B\rightarrow D^\pm X$, respectively. This has to be compared with the decay
length the charm mesons of $c\tau=124~\mu m$ for the $D^0/\overline D^0$ and
of $c\tau=317~\mu m$ for the $D^\pm$. Requiring a selection efficiency for the
cut in the impact parameter of 50\% would reduce the contamination of charm
events by a factor of 12.5(2.8) in case of the $D^0(D^\pm)$ analysis. 
\end{enumerate}
All decay modes summarized in Tab.~\ref{tab1} will also be
 accessible in this case.
Due to the larger cut values in the impact parameter
for the different decay modes the combinatorial
background is smaller. This will enable us to use softer cuts. Compared to 
the measurement of the charm production cross section in total 
an increase in efficiency
by a factor 3 may be expected. A systematic error of 16\% may be achieved. It
will be dominated by the uncertainties in the branching fraction 
$BR(B\rightarrow DX)$.  
\item the analysis of the impact parameter distribution of at least 3 to 4 
tracks fitted to a common secondary vertex. Compared to the exclusive analysis
this method leads to a higher selection efficiency, but will be limited 
by larger systematic uncertainties in the subtraction of the charm quark 
induced background.
\end{enumerate}

\begin{figure}[t]
\begin{picture}(17.8,6.3)
\put(1.0,-1.7){\epsfig{file=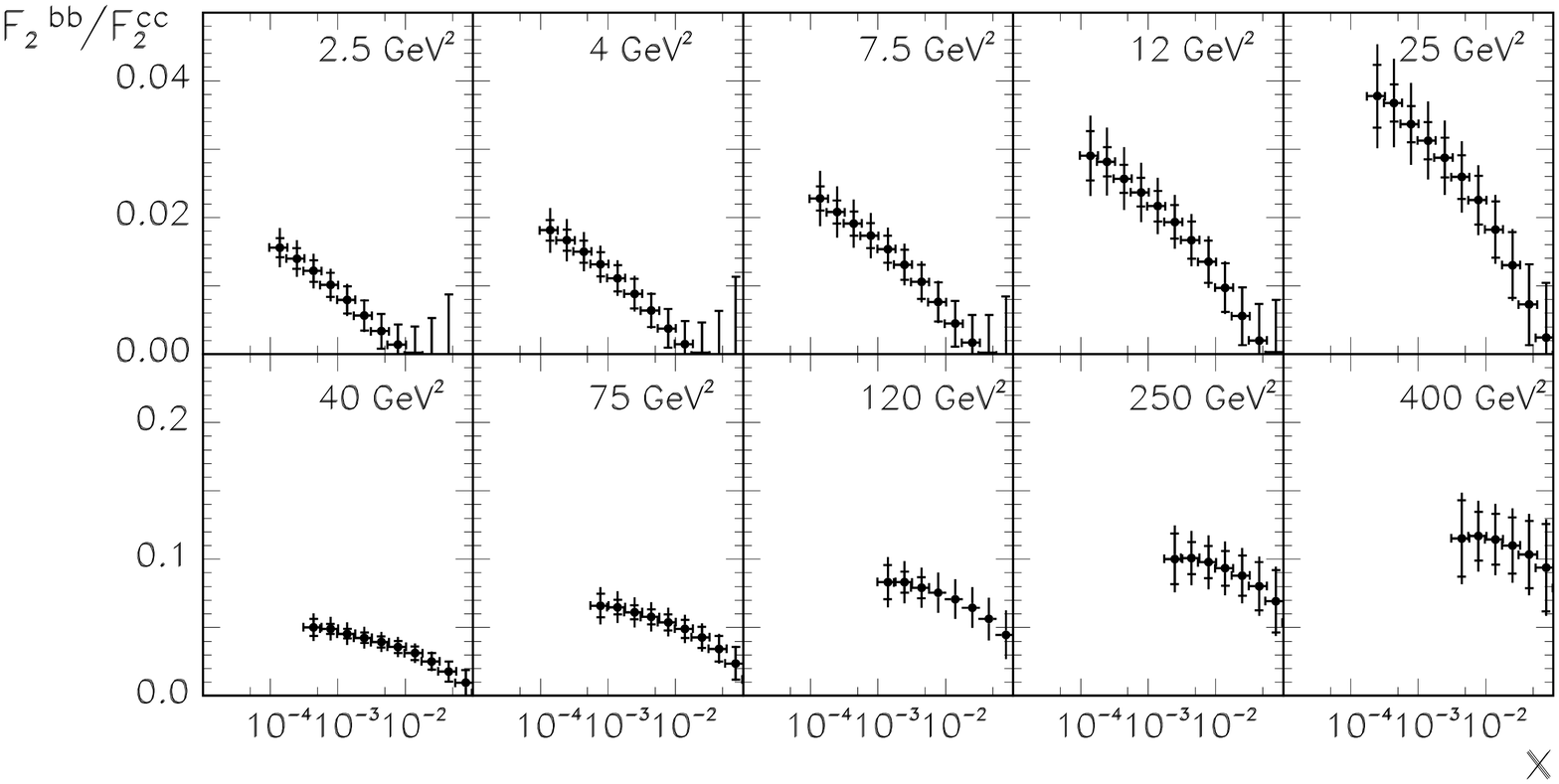,width=15cm}}
\end{picture}
\centering
\caption{ {\it Expected ratio $F_2^{b\overline b}/F_2^{c\overline c}$ 
          for a luminosity of 500~$pb^{-1}$. 
          The points show the predictions according to the calculations of 
          section 2.1 and 2.6. The inner (outer) error 
          represents the statistical and the total experimental error. 
          }}
\label{f2bf2c}
\end{figure}
Figure \ref{f2bf2c} shows the ratio of $F_2^{b\overline b}/F_2^{c\overline c}$
expected for the exclusive $D$ meson analysis from bottom quark
production using the the calculations of sections 2.1 and 2.6. The charm
production background is subtracted statistically.  The errors 
refer to the statistical and the total experimental error. For a luminosity of
500~pb$^{-1}$ the error on the measurement of this ratio will still be 
statistics limited in most of the $x$ and $Q^2$ plane. For a given
$Q^2$ or $x$, 
this ratio is expected to rise with decreasing $x$ or increasing $Q^2$, 
respectively. Integrated over the kinematic range a mean value of
$F_2^{b\overline b}/F_2^{c\overline c}\approx 0.02$ is predicted.

\section{Conclusions}

In this paper, we have discussed DIS heavy-flavour production at HERA
at NLO. The effect on the charm production cross section due to the 
uncertainties in the factorization/renormalization scale $\mu$ and the 
charm quark mass $m_c$ has been studied as a function of $x$ and $Q^2$. At 
small $x$ and for $Q^2<300~\mbox{GeV}^2$ the predictions are found to be 
insensitive to $\mu$ while in this region the effect of $m_c$ is found to be 
large. Unfortunately the sensitivity of  
$F_2^{c\overline c}(x,Q^2)$ to the gluon density is also restricted to this 
kinematic region. It has been shown that the contribution of $F_L$ has 
sizeable effects on the charm production cross section at large $y$. 
The contribution of light quarks to the cross section turned out to be of the
order of 5\%, nearly independent of $x$ and $Q^2$. Results on bottom 
production have been presented.

The current experimental situation has been summarized. Based upon
this knowledge,
statistical and experimental systematic uncertainties for large 
luminosity have been estimated. Ultimately an accuracy of 10\% in the overall
normalization of the cross section may be achieved. In the kinematic range 
where the inclusive measurement  $F_2^{c\overline c}(x,Q^2)$ is found to be 
sensitive to the gluon density, the error of the inclusive measurement will 
start to be systematics dominated with a luminosity of 50~$pb^{-1}$. It has
been demonstrated that exclusive measurements of the charm mesons would
disentangle the influence of $m_c$ and the gluon density on the 
charm production cross section.  An exploration of the kinematic plane,
the extraction of the gluon density, as well as the question when the
charm quark may be treated as a parton, will require a luminosity of
500~$pb^{-1}$. Finally the ratio of  
$F_2^{b\overline b}/F_2^{c\overline c}(x,Q^2)$ has been investigated by
performing an exclusive $D$ meson analysis. With a luminosity of
500~$pb^{-1}$ the predicted statistics is found
to be sufficient to make a detailed study of the $x$ and $Q^2$ dependence
of this ratio.

\noindent
{\bf Acknowledgments}

S.R. thanks Jenny Ivarsson for her help making the figures and Fred
Jegerlehner for reading the manuscript.

\end{document}